\documentclass[12pt,a4paper]{article}
\usepackage{amssymb,amsfonts,amsmath}

\begin{document}

\title{Ozawa's Intersubjectivity Theorem as objection to QBism individual agent perspective}

\author{Andrei Khrennikov\\ 
Linnaeus University, International Center for Mathematical Modeling\\  in Physics and Cognitive Sciences
 V\"axj\"o, SE-351 95, Sweden}

\maketitle

\abstract{QBism's  foundational statement that ``the outcome of a measurement of an observable is personal''  is in the straight contraversion with Ozawa's Intersubjectivity Theorem (OIT). The latter (proven within the quantum formalism) states that two observers, agents within the  QBism terminology, performing joint measurements of the same observable $A$ on a system $S$ in the state $\psi$ should get the same outcome   $A=x.$ In Ozawa's terminology, this outcome is intersubjective and it can't be treated as personal. This is the strong objection to QBism  which can't survive without updating its  principles. The essential aspect in understanding  of the OIT-impact on QBism's foundations takes the notion of quantum observable. This paper comprises the complementary discussion highlighting the difference between the accurate, von Neumann, and inaccurate, noisy, quantum observables which are represented by PVMs and POVMs respectively. Moreover, we discuss the OIT-impact  on the Copenhagen interpretation of quantum mechanics.}     

\section{Introduction}
In this paper I move ahead my critical analysis of QBism's foundations (see, e.g., \cite{Fuchs3}--\cite{Fuchs7} for QBism basics). This paper, as well as my two previous articles \cite{KHRQBism1,KHRQBism2}, straightly critiques the individual agent perspective on measurement's outcomes \cite{KHRBrukner}. My previous appraisal convinced QBists to specify the level of agent's individuality. In contrast to the general subjective probability theory, the class of agents should be restricted, at least to agents who were educated in basics of quantum theory. So, Ivan who lives in a Siberian village, a busy hunter, can't be treated as a QBism's agent. 

Now I have an intention to offense QBism by using  Ozawa's Intersubjectivity Theorem (OIT) \cite{OIT}. Qbism's statement that ``the outcome of a measurement of an observable is personal'' is in the straight contraversion with OIT.  This theorem is not so widely known and one of the present paper's intention is the theorem's advertizement. OIT states that two observers, agents within the QBism terminology, performing joint measurements of the same observable $A$ on a system $S$ in the state $\psi$ should register the same outcome  $A=x$ with probability one. Hence, the outcome is intersubjective \cite{OIT}, and it's unnatural to consider outcomes of quantum observations as agent's personal experiences. 

OIT is proven within the  quantum formalism, it is the rigorous  mathematical statement. But, as many theorems having the quantum foundational impact, its interpretation is not straightforward. The analysis of the OIT-impact onto QBism is coupled to 
the foundations of quantum measurement theory and especially the notion of quantum observable. Therefore, this paper comprises the complementary discussion, highlighting the difference between the accurate, von Neuman, and inaccurate, noisy, quantum observables, mathematically represented by projection valued measures (PVMs) and  positive operator valued measures  (POVMs), respectively.  
QIT is about the agents who are able to perform the joint accurate measurements. For such agents, measurement's outcome loses its personalization, in favour of intersubjectivity.   

The conclusion of our analysis is that QBism should update its ideology by taking in consideration  OIT. But, how? See section \ref{Conclusion}. Thus, I am in line with the criticism of QBism presented in article \cite{OIT}. However, I depart from its conclusion that OIT contradicts to the Copenhagen interpretation; in contrast, OIT  peacefully coexist with this interpretation. It is relevant to recall here that QBism  fundamentally differs  from the Copenhagen interpretation \cite{Fuchs5}.   

Right away we initiate with the mathematical formulation of OIT and its proof. We set out to make the presentation very shortly (see \cite{OIT} for details). The indirect measurement scheme is the heart of OIT. We go ahead with the recollection of the notion of quantum observable, namely, Hermitian operator or PVM, and generalized quantum observable (POVM) and the indirect measurements  scheme for their generation.

\section{Quantum observables vs. generalized  quantum observables}

In quantum mechanics' axiomatics, von Neumann \cite{VN} introduced quantum observables as Hermitian operators acting in complex Hilbert space ${\cal H},$ the state space of a system.\footnote{Why did he select the Hermitian operators for mathematical representation of observables in quantum theory? Moreover, he considered only such observables as the genuine quantum observables. I guess that he followed Schr\"odinger's quantization rule for the position and momentum observables which are realized by Hermitian operators in $L_2$-space. This rule implies that each classical observable given by the real-valued function $A=A(q, p)$ on the phase space is represented as a Hermitian operator in $L_2$-space.}  The spectral decomposition is the essential part in this framework.  

We restrict considerations to observables represented by the operators with totally discrete spectra $X \subset 
\mathbb R.$ Here  
\begin{equation}
\label{MA5}
A=\sum_{x} x E_A(x),
\end{equation} 
where $E_A(x)$ is projection on the eigensubspace corresponding  to the eigenvalue $x;$ these projectors form the resolution of unity:  
\begin{equation}
\label{MA6}
I =\sum_x  E_A(x).
\end{equation} 
The Born rule determines the probabilities of the outcomes of measurements for a system ${\cal S}$ in the state $\psi,$ 
\begin{equation}
\label{MA7}
P(A = x| \psi) = \langle \psi | E_A(x) |\psi \rangle.
\end{equation} 

Later generalized quantum observables were invented. Such observables are represented by POVMs. We restrict considerations to  POVMs with a discrete domain of definition $X.$  POVM is a map $x \to  \Pi(x):$ for each $x \in X,  \Pi(x)$ is a positive contractive
self-adjoint operator (i.e.,  $0\le \Pi(x)\le I$) (called an {\it effect}), and  effects form the resolution of unity 
\begin{equation}
\label{zzz}  
\sum_x \Pi(x)=I .
\end{equation}
This map defines an operator valued measure on algebra of all subsets of set $X.$
For $O \subset X,$ $$\Pi(O)= \sum_{x \in O} \Pi(x).$$ The condition (\ref{zzz}) 
is the  operator-measure counterpart of the condition normalization by 1 for usual probability measures.

POVM  $\Pi$ represents statistics of  measurements for observable $A$ with the following generalization  of the Born's rule:   
\begin{equation}
\label{z1}  
P(\Pi =x  |\psi) =  \langle \psi | \Pi(x) |\psi \rangle . 
\end{equation}
We remark that equality (\ref{zzz}) implies that $$\sum_x P(A=x  |\psi)=1.$$
Any quantum observable $A$ can also be represented as POVM of the special type -- PVM 
$E_A=(E_A(x)).$

Quantum observables given by PVMs were interpreted by von Neumann \cite{VN} as describing {\it accurate measurements}. And generalized observables given by POVMs which are not PVMs are interpreted as representing {\it inaccurate measurements}. In von Neumann's \cite{VN}, the notion of measurement's precision was not completely formalized. Only recently the consistent formalization of this notion was presented in \cite{O4}.

We shall keep firmly the expression ``quantum observable'' for observable axiomatically introduced by von Neumann \cite{VN} and represented by PVMs and the expression ``generalized quantum observable'' for POVMs.

\section{Generalized quantum observables from the indirect measurement scheme}

The indirect measurement scheme involves the following components
\begin{itemize}
\item the states spaces ${\cal H}$ and ${\cal K}$ of the systems ${\cal S}$ and the apparatus ${\cal M}$ for measurement of some observable $A;$     
\item the evolution operator $U=U(t)$ representing the interaction-dynamics for the system ${\cal S}+ {\cal M};$ 
\item the meter observable $M$  giving outputs of the pointer of the apparatus ${\cal M}.$ 
\end{itemize}
Here  the quantum observables $A$ and $M$ can be represented as  PVMs, $E_A=(E_A(x)), E_M=(E_M(x)),$ where $E_A(x), E_M(x)$ are  projections in  Hilbert spaces ${\cal H}$ and ${\cal K}$ respectively. It is assumed that the compound system's  evolution is driven by the Schr\"odinger equation, so the  evolution operator is unitary. 

Formally, an {\em indirect measurement model} for an observable $A$, introduced in \cite{O1} as a ``measuring process'', 
is a quadruple 
$$
({\cal K}, |\xi\rangle , U, M)
$$ 
where  $|\xi\rangle \in {\cal K}$ represents the apparatus state.  

We explore the Heisenberg picture. To describe meter's evolution, we represent it in the state space of the compound system, i.e., as
$I \otimes M.$  The meter observable evolves as
\begin{equation}
\label{MA1}
M(t) = U^\star(t)(I\otimes M) U(t).
\end{equation} 
By the Born rule 
\begin{equation}
\label{MA2}
P(M(t)= x| \psi \xi) = \langle \psi \xi | E_{M(t)}(x)| \psi \xi \rangle .
\end{equation} 

This is the probability distribution for the outputs of measurements done by the apparatus and given by the meter.
In principle, one can ignore the representation of the measurement process as the system-apparatus interaction and operate 
solely with system's states. In this picture one proceeds with generalized observables given by POVMs. The meter observable generates 
the POVM   $\Pi=(\Pi(x))$
\begin{equation}
\label{MA3}
\Pi(x)=  \langle \xi | E_{M(T)}(x)|  \xi \rangle,
\end{equation} 
where $T$ is the time needed to complete the experiment.

The probability distribution of the generalized observable given by a POVM is determined by (\ref{z1}). 

Generally the probability distribution generated  by a measurement process does not  coincide with the probability distribution 
of the quantum observable $A$ for which this process was constructed, i.e., generally
\begin{equation}
\label{MA4}
P(\Pi = x| \psi) = \langle \psi | \Pi(x) |\psi \rangle \not= P(A = x| \psi) = \langle \psi | E_A(x) |\psi \rangle,
\end{equation} 

We remark that, as was proven by Ozawa \cite{O1}, any generalized observable (POVM) can be generated via the indirect measurement scheme.
Typically one operates solely with generalized observables by ignoring the indirect measurement scheme. This simplifies considerations, but it can lead to misunderstanding of the foundations the quantum measurement theory.

\section{Probability reproducibility condition}

{\bf Definition.} {\it A measurement process $({\cal K}, |\xi\rangle , U, M)$ reproduces the probability distribution for  quantum observable $A$  (accurate von Neumann observable)  if} 
\begin{equation}
\label{MA7}
P(A = x| \psi) = P(M(T)= x| \psi \xi).
\end{equation} 

\medskip

In this case
\begin{equation}
\label{MA7t}
 \langle \psi \xi | E_{M(T)}(x)| \psi \xi \rangle =\langle \psi | E(x) |\psi \rangle.
\end{equation} 
or 
\begin{equation}
\label{MA7q}
 \langle \psi | \Pi(x)| \psi \rangle =\langle \psi | E(x) |\psi \rangle,
\end{equation} 
and hence, 
$$
\Pi(x)= E(x),
$$

{\bf Proposition.} {\it Probability reproducibility condition for a measurement process is equivalent to the representation 
of the corresponding generalized observable by the PVM $E_A$ of measured quantum observable $A.$} 

\section{Intersubjectivity of outcomes of quantum observables}

Following \cite{OIT}, consider two remote observers $O_1$ and $O_2$ who perform joint measurements on a system ${\cal S},$  in mathematical terms it means that the meter quantum observables of the corresponding measurement processes commute, 
 $$
[M_1(t), M_2(t)]=0.
$$ 
Here each apparatus has its own state space, i.e., 
${\cal K}= {\cal K}_1 \otimes {\cal K}_2.$ We call such measurements local. 
In this situation the joint probability distribution is well defined 
\begin{equation}
\label{MA9}
P(M_1(t)= x, M_1(t)= y | \psi \xi_1 \xi_2) = \langle \psi \xi_1 \xi_2| E_{M_1(t)}(x) E_{M_1(t)}(y)| \psi \xi_1 \xi_2 \rangle
\end{equation}

Suppose that both observers perform the accurate measurements of the quantum  observable $A$ given by PVM $E_A=(E_A(x)).$ Then the corresponding POVMs $\Pi_j, j=1,2,$ coincide with $E_A:$
\begin{equation}
\label{MA9}
\Pi_1(x)= \Pi_2(x)= E_A(x).
\end{equation}
This equality implies: 

\medskip

{\bf Theorem.} (OIT \cite{OIT}) {\it Two observers performing the joint local and probability reproducible  measurements of the same quantum observable $A$  on the system ${\cal S}$ should get the same outcome with probability 1:}
\begin{equation}
\label{MA10}
 P(M_1(T)= x, M_1(T)= y | \psi \xi_1 \xi_2) = \delta (x-y) P (E= x | \psi) = \Vert E(x) \psi \Vert^2.
 \end{equation}

\section{Intersubjectivity challenges QBism}
\label{Conclusion}

 We start with the following citation of  Fuchs and Schack \cite{Fuchs5}:

{\it ``The fundamental primitive of QBism is the concept of experience. According to QBism,
quantum mechanics is a theory that any agent can use to evaluate her expectations for
the content of her personal experience. ...

In QBism, a measurement is an action an agent takes to elicit an experience.
The measurement outcome is the experience so elicited. The measurement outcome
is thus personal to the agent who takes the measurement action. In this sense, quantum
mechanics, like probability theory, is a single user theory. A measurement does not reveal
a pre-existing value. Rather, the measurement outcome is created in the measurement
action.''} 

However, OIT implies that, for accurate local observables, measurement's outcome is  intersubjective  which is the strong objection to QBism.  There is nothing concerning personal experiences and QBists should response to this objection. My suggestion (see also  \cite{KHRBrukner}) is to follow Brukner's work \cite{Brukner} where he proceeds not with individual agents  and their personal experiences, but with a {\it universal agent.}  I remark that consideration of universal agents is common in general theory of decision making. However, for QBists, such solution seems to be unacceptable, since it would destroy consistency of the QBism's private agency perspective. It would move QBism closer to Zeilinger-Brukner information  interpretation of quantum mechanics \cite{Z,BZ1,BZ2}.

This objection to QBism is foundationally interesting and generates the discussion on the notion of quantum observable. Due to efforts  Helstrom, Holevo, and Ozawa \cite{Hel76}--\cite{89RS}, \cite{O1}, generalized quantum observables which are mathematically represented by POVMs became one of the basic tools of quantum information theory.  Nowadays the special role of accurate observables represented by PVMs is not emphasized.  In particular,  the notion of observables  in QBism is identified with  generalized  quantum observable given by POVM. However, the clash between QBism and  OIT stimulates highlighting of the accurate PVM- as the genuine quantum observables, and treating the generalized quantum observables which are not accurate POVM as imprecise and noisy ones. Of course, it is a well known fact, but the clash  between OIT and QBism is good occasion to emphasize this difference.

{\it What does this difference between accurate PVM and noisy POVM observables mean for QBism?}

I have the following picture of the situation. OIT holds only for the accurate PVM-observables; for generalized quantum observables, it can be violated  and generally it is impossible 
to assign the same value for measurements' outcomes for observers $O_1$ and $O_2.$ 
Thus, QBism ideology of the personal experiences of observers (agents) can still be kept for such generalizad observables.  But, where does individuality come from? {\it The personal experiences  come from noise!} So, different observers performing inaccurate measurements are coupled to different noisy environments. This is just my personal view on consequences of IOT for QBism.

In conclusion, QBism might response to the OIT-challenge by considering the universal agent who is able to perform accurate measurements; individuality of agents' experience is reduced to individuality of noise generated in the process of measurement.    
 
\section{Intersubjectivity and Copenhagen interpretation}

By the Copenhagen interpretation (at least by its Bohr's version\footnote{As was stressed by Plotnitsky \cite{PL2}, one should recognize the diversity of views on  the Copenhagen interpretation. He suggested to speak about interpretations {\it in the spirit of Copenhagen.} Even Bohr changed  the views a few times during his life \cite{PL2}.}) measurements' outcomes cannot be treated as the objective properties 
of a system ${\cal S}.$ They are results of the complex process of interaction of a system and an apparatus, see 
Bohr \cite{BR0}:

\medskip

{\it ``This crucial point ...  implies the impossibility of any sharp separation between the behaviour of atomic objects and the interaction with the measuring instruments which serve to define the conditions under which the phenomena appear. In fact, the individuality of the typical quantum effects finds its proper expression in the circumstance that any attempt of subdividing the phenomena will demand a change in the experimental arrangement introducing new possibilities of interaction between objects and measuring instruments which in principle cannot be controlled. Consequently, evidence obtained under different experimental conditions cannot be comprehended within a single picture, but must be regarded as complementary in the sense that only the totality of the phenomena exhausts the possible information about the objects.''} 

\medskip

The indirect measurement scheme matches perfectly with the Copenhagen interpretation. Therefore it is surprising that OIT contradicts to it. The clash between OIT and the the Copenhagen interpretation was highlighted in  the conclusion section of OIT-article \cite{OIT}:

\medskip

{\it ``Schr\"odinger \cite{present} argued that a measurement does not
ascertain the pre-existing value of the observable and is only
required to be repeatable. Since the inception of quantum mechanics,
this view has long been supported as one of the fundamental
tenets of quantum mechanics. In contrast, we have
shown that any probability reproducible measurement indeed
ascertains the value that the observable has, whether the repeatability
is satisfied or not.''}
   
\medskip

I disagree with the author of  \cite{OIT}. The seed of  this misunderstanding is in ignoring the two  level structure of physical theories, ontic and epistemic \cite{ATM0,ATM,KHRontic}. The former is about reality as it is and the latter is about knowledge about reality. Bohr and Schr\"odinger wrote about the ontic reality, about impossibility to assign  to quantum systems  preexisting values and here ``preexisting'' is the synonym for  ``objective'', ``ontic''. But OIT is not about such values,  it is about epistemic reality, reality of knowledge about the possible outcome of measurement.  

Hence, in my opinion {\it OIT can peacefully coexist with the Copenhagen interpretation.}

But, as was stressed, OIT is a challenge for QBism which operates at the epistemic level of 
scientific description of quantum phenomena.  This is the good place to recall that QBism should be sharply separated from the Copenhagen interpretation, see again Fuchs and Schack \cite{Fuchs5}:

{\it ``According to QBism, quantum mechanics can be applied to any physical system.
QBism treats all physical systems in the same way, including atoms, beam splitters,
Stern-Gerlach magnets, preparation devices, measurement apparatuses, all the way to
living beings and other agents. In this, QBism differs crucially from various versions
of the Copenhagen interpretation.''}

\section*{Acknowledgments} 

This paper was written on the basis of the long discussions with Masanao Ozawa and I would like to thank him; Arkady Plotnitsky told me a lot about the Copenhagen interpretation and Bohr's views and I would like  to thank him;  Christopher Fuchs ignited my interest to QBism at  the second V\"axj\"o conference (in 2001) and I am sorry if this paper would disturb  QBists; I am also thankful to Harald Atmanspacher who introduced me into ontic-epistemic approach to scientific representation of natural phenomena.

\end{document}